\documentclass[11pt]{article}

\usepackage[T1]{fontenc}
\usepackage[utf8]{inputenc}

\usepackage{fancyhdr}

\usepackage[inline]{enumitem}
\usepackage{multicol,multirow,booktabs}

\usepackage{times}

\usepackage[hang,small,labelfont=bf,up,textfont=it,up]{caption}
\usepackage{subcaption}
\usepackage{amsmath,amsfonts,amsthm,amssymb,bm,dsfont}	
\setlist{nolistsep} 

\usepackage{graphicx}
\usepackage[a4paper,top=2.75cm,bottom=2.75cm,left=1.75cm,right=1.75cm]{geometry}

\usepackage{numprint}
\npthousandsep{,}\npthousandthpartsep{}\npdecimalsign{.}

\newcommand{\titledoc}{Graph-based mutually exciting point processes for modelling event times in docked bike-sharing systems}
\newcommand{\titleshort}{Graph-based mutually exciting point processes for modelling event times in docked BSSs}

\providecommand{\keywords}[1]{{\small{\textbf{\textit{Keywords ---}} #1}}}

\usepackage{fontawesome}
\usepackage{natbib}
\usepackage{setspace}

\usepackage{thmtools}
\newtheorem{proposition}{Proposition}

\usepackage{authblk}

\usepackage{mathtools,bigints}
\mathtoolsset{showonlyrefs}

\usepackage[ruled,linesnumbered,noend]{algorithm2e}
\SetKwInput{KwInput}{Input}
\let\oldnl\nl
\newcommand{\nonl}{\renewcommand{\nl}{\let\nl\oldnl}}
\makeatother

\author{Francesco Sanna Passino}
\author{Yining Che}
\author{Carlos Cardoso Correia Perello}

\affil{Department of Mathematics, Imperial College London \\ 180 Queen's Gate, SW7 2AZ, London}

\date{}

\title{\huge\textbf{\titledoc}}

\allowdisplaybreaks

\usepackage[colorlinks=true,linkcolor=blue,citecolor=blue,urlcolor=blue]{hyperref}
\numberwithin{equation}{section}

\begin{document}

\maketitle


\renewcommand\thefootnote{}
\footnotetext{
\noindent For the purpose of open access, the authors have applied
a Creative Commons Attribution (CC-BY) licence to any
\textit{Author Accepted Manuscript} version arising. \\
CCCP completed this work at Imperial College London, and he is currently in the Department of Mathematics at the University of Oxford.  \\
Corresponding author: Francesco Sanna Passino -- \faEnvelopeO\ \texttt{f.sannapassino@imperial.ac.uk}
}

\begin{abstract}
This paper introduces graph-based mutually exciting processes (GB-MEP) to model event times in network point processes, focusing on an application to docked bike-sharing systems. GB-MEP incorporates known relationships between nodes in a graph within the intensity function of a node-based multivariate Hawkes process. This approach reduces the number of parameters to a quantity proportional to the number of nodes in the network, resulting in significant advantages for computational scalability when compared to traditional methods. 
The model is applied on event data observed on the Santander Cycles network in central London, demonstrating that exploiting network-wide information related to geographical location of the stations is beneficial to improve the performance of node-based models for applications in bike-sharing systems. The proposed GB-MEP framework is more generally applicable to any network point process where a distance function between nodes is available, demonstrating wider applicability. 
\end{abstract}


\section{Introduction} \label{sec:intro}

In recent years, a number of convenient and sustainable alternatives for short-distance travel have emerged in world cities, with the objective of reducing traffic congestion, and promoting healthier and more sustainable urban environments. Bike-sharing systems (BSSs) represent one such solution, which have become increasingly popular especially after the end of restrictions to movement due to the COVID-19 pandemic \citep{Heydari21}. Two types of BSSs are usually offered in cities: traditional docked systems, which require users to pick up and return bikes at designated stations, and dockless systems, which utilise GPS technology, allowing users to locate and unlock bikes using smartphone applications. From a statistical perspective, the literature about BSSs is mostly focused on two problems: rebalancing, consisting in the redistribution of bikes across stations and locations according to the predicted demand \citep[see, for example, the survey of][]{Vallez21}, and demand forecasting \citep[for example,][]{Rennie23}. In general, modelling docked bike-sharing systems is still an active research problem, extensively studied in the literature \citep[see, for example,][]{Choi22,Deng23,Busatto23,Paul23}. 

Motivated by the application to docked bike-sharing systems, this work proposes a statistical network-informed model for event times, generically called \textit{graph-based mutually exciting process} (GB-MEP), which could be used to predict demand across stations. The proposed methodology is tested on publicly available data from the Santanter Cycles bike-sharing network in central London.
In this work, journeys observed on bike-sharing systems are interpreted as a realisation of a network point process, with events having a corresponding duration. In general, demand for bicycles in docked BSSs has often been modelled as a multivariate point process in the literature \citep[see, for example,][]{Gervini19,Okawa19,Foschi23}, but network information, expressing the interaction between different stations, is often not explicitly incorporated within the existing models. This work proposes a parsimonious representation for the temporal point processes observed at each docking station, directly utilising the information provided by the distance between stations in the network for imposing restrictions on the model parameters. 

In particular, this work considers a particular class of point processes, called mutually exciting processes, with a focus on Hawkes processes \citep[see, for example,][]{Hawkes71,Laub22}.  
Hawkes processes have been extensively used in the literature for modelling a number of different phenomena, such as earthquakes \citep{Ogata88}, crime \citep{Mohler11}, social media \citep{Zhao15, Rizoiu17}, emails \citep{Fox16, Miscouridou18,SannaPassino23}, and computer networks \citep{PriceWilliams19, Shlomo22, SannaPassino23}.
Hawkes processes appear to be a particularly suitable model for a bike-sharing system, since they capture periods of increased demand in a data-driven way, avoiding specific parametric assumptions on the process, such as periodic components or weather conditions. This property has been proved particularly powerful in cyber-security, where self-exciting models largely outperform alternatives that model periodic components parametrically \citep{PriceWilliams19}. 

The remainder of this article is structured as follows: Section~\ref{sec:background} gives background on point processes on graphs and Hawkes processes, describing related literature and challenges of existing approaches when applied to network point processes. Next, Section~\ref{sec:gbmepp} introduces graph-based mutually exciting processes, the main contribution of this work, discussing the method, parameter estimation, and model evaluation procedures. The proposed methodology is then applied on publicly available data from the Santander Cycles bike-sharing system in central London (Section~\ref{sec:results}), followed by a conclusion and discussion in Section~\ref{sec:discussion}.

\section{Background} \label{sec:background}

Consider a network $\mathbb G=(\mathcal V,\mathcal E)$ where $\mathcal V=\{1,\dots,M\},\ M\in\mathbb N$, is the node set, and $\mathcal E\subseteq \mathcal V\times\mathcal V$ is the edge set, such that $(x,y)\in\mathcal E$ if nodes $x$ and $y$ are connected. Within the context of bike-sharing systems, nodes represent docking stations, and an edge between two stations can be drawn if at least one trip between the stations is observed within an observation period. It is further assumed that event data with a duration are observed on the network, corresponding to a sequence of quadruples $(x_k, y_k, t_k, t_k^\prime),\ k=1,2,\dots,N,\ N\in\mathbb N$, where $(x_k,y_k)\in\mathcal E$ is an edge, $t_k\in\mathbb R_+$ is the start time of the event, and $t_k^\prime>t_k$ is the end time of the event. This corresponds to observing events with duration $\delta_k=t_k^\prime-t_k$. In practice, events are ordered by end time, such that $t_1^\prime<\cdots<t_N^\prime$. In bike-sharing systems, the quadruple $(x_k, y_k, t_k, t_k^\prime)$ denotes a bike journey between stations $x_k$ and $y_k$, started at time $t_k$ and ended at time $t_k^\prime$. 
From the observed sequence of quadruples, two left-continuous counting processes $N_i:\mathbb R_+\to\mathbb N_0$ and $N_i^\prime:\mathbb R_+\to\mathbb N_0$ could be associated with each node:  
\begin{align}
N_i(t) = \sum_{k \geq 1} \mathds{1}_{[0,t)}{(t_k)} \mathds{1}_{\{i\}}{(x_k)}, & &
N^\prime_i(t) = \sum_{k \geq 1} \mathds{1}_{[0,t)}{(t_k^\prime)} \mathds{1}_{\{i\}}{(y_k)}, \label{eq:counting}
\end{align}
where $\mathds 1_\cdot(\cdot)$ denotes the indicator function: $\mathds 1_S(x)=1$ if $x\in S$, and $0$ otherwise.
In the BSS application, the process $N_i(t)$ counts the number of journeys started from station $i$ before time $t$, whereas the process $N^\prime_i(t)$ counts the number of journeys ended at station $i$ before time $t$. For simplicity, event times for each node are often denoted with the notation $t_{i,k},\ i\in\mathcal V,\ k\in\mathbb N$, representing the start time of the $k$-th event started from node $i$, and $t^\prime_{i,k}$, corresponding to the ending time of the $k$-th event ended on node $i$. This notation can be used to obtain a simpler representation for $N_i(t)$ and $N_i^\prime(t)$: 
\begin{align}
N_i(t) = \sum_{k \geq 1} \mathds{1}_{[0,t)}{(t_{i,k})}, & & 
N^\prime_i(t) = \sum_{k \geq 1} \mathds{1}_{[0,t)}{(t_{i,k}^\prime)}. \label{eq:counting_simple}
\end{align}
In general, a counting process could be characterised by its conditional intensity function (CIF). The CIF $\lambda_i(t)$ for $N_i(t)$ is defined as follows:
\begin{align}
\lambda_i(t) = \lim_{\mathrm dt\to0}\frac{\mathbb E[N_i(t+\mathrm dt) - N_i(t) \vert\mathcal H(t)]}{\mathrm dt}, \label{cif}
\end{align}
where $\mathcal H(t)$ is the history of the process on the entire network up to time $t$. 

The counting processes $\{N_i(t)\}_{i\in\mathcal V}$ and $\{N_i^\prime(t)\}_{i\in\mathcal V}$ could be considered as realisations of multivariate temporal point processes. For this class of processes, a popular model in the literature is the multivariate Hawkes process \citep{Hawkes71} with scaled exponential kernel \citep[see, for example,][]{Laub22}, which models the CIF $\lambda_i(t)$ on each node as a function of all start times of events observed in the entire network before time $t$:
\begin{equation}
\lambda_i(t) = \lambda_i + \sum_{j=1}^M \sum_{t_{j,k} < t} \alpha_{ij}\exp\{-\beta_{ij}(t-t_{j,k})\}, \label{eq:cif_hawkes}
\end{equation}
where $\lambda_i\in\mathbb R+$ is a baseline intensity, $\alpha_{ij}\in\mathbb R_+$ is the jump in the CIF of node $i$ caused by an event on node $j$, and $\beta_{ij}\in\mathbb R_+$, with $\alpha_{ij}<\beta_{ij}$, is the corresponding decay rate. The end times of events could also be considered in the intensity in Equation~\eqref{eq:cif_hawkes}, obtaining the following self-and-mutually exciting Hawkes process CIF \citep[see, for example,][]{Laub22}:  
\begin{equation}
\lambda_i(t) = \lambda_i + \sum_{j=1}^M \left[ \sum_{t_{j,k} < t} \alpha_{ij}\exp\{-\beta_{ij}(t-t_{j,k})\} + \sum_{t_{j,h^\prime} < t} \alpha_{ij}^\prime\exp\{-\beta_{ij}^\prime(t-t_{j,h}^\prime)\}\right], \label{eq:cif_hawkes_mut}
\end{equation}
where $\alpha_{ij}^\prime,\beta_{ij}^\prime\in\mathbb R_+$, $\alpha_{ij}^\prime < \beta_{ij}^\prime$. 
The models in Equations \eqref{eq:cif_hawkes} and \eqref{eq:cif_hawkes_mut} have a number of issues when applied to temporal point processes on networks: first, the number of parameters is $\mathcal O(M^2)$ across the entire network, implying that it grows quadratically with the number of nodes, which is unfeasible for estimation in practice, especially when the network becomes large; second, the network structure is not explicitly incorporated within the intensity function in Equations~\eqref{eq:cif_hawkes} and \eqref{eq:cif_hawkes_mut}; third, information from additional processes, such as end times of events, is not included within the CIF in Equation~\ref{eq:cif_hawkes}. 

In the application to bike-sharing systems, these issues appear to be particularly relevant: the number of docking station could be very large (approximately $800$ in the Santander Cycles BSS in central London), making estimation of the parameters $\alpha_{ij}$ and $\beta_{ij}$ a high-dimensional problem. Using the example of $M=800$, the total number of parameters required for model in Equation~\eqref{eq:cif_hawkes} for the entire network would be $M+2M^2=\numprint{1280800}$. This number is more than the average number of bike journeys per month in the busiest year in the Santander Cycles BSS \citep[over $\numprint{950000}$ hires per month in 2022, see][]{TFL22}. Additionally, another drawback of the multivariate Hawkes process models in Equation~\eqref{eq:cif_hawkes} and \eqref{eq:cif_hawkes_mut} is that information about \textit{known}, \textit{pre-existing} network structure is not incorporated within the CIF: for modelling network point processes in real-world applications, information about the nodes in the BSS could be very important. For example, in a bike-sharing system, stations which are geographically close are expected to have similar characteristics compared to stations which are further apart, suggesting that geographical information about nodes should be taken into account within the model. 

It must be remarked that the model in Equation~\eqref{eq:cif_hawkes} has been often used successfully in the literature for modelling multivariate point processes, including spatio-temporal processes \citep[see, for example, the survey of][]{Reinhart18}. In particular, Hawkes processes are in general suitable to \textit{discover} existing network structure and causal relationships between nodes \citep[see, for example,][]{Linderman14, Etesami16, Xu16, Eichler17, Yuan19}, but known characteristics about the network and nodes are not explicitly incorporated within the model. Using the approach of \cite{Linderman14}, the network adjacency matrix, if known, could be used to set the coefficients $\alpha_{ij}$ to zero where $(i,j)\not\in\mathcal E$, but in general the use of node covariates and known network information has been overlooked in the literature. Also, it must be remarked that Hawkes process models on networks are usually applied at two different levels of resolution: node-based \citep[see, for example,][]{Fox16,PriceWilliams19}, or edge-based \citep[see, for example][]{Blundell12,Perry13,Miscouridou18,Huang22,SannaPassino23}. 

This work mostly concerns with node-based models, proposing an extension of the standard multivariate Hawkes process, exploiting existing network structure between the nodes to reduce the number of parameters to estimate to $\mathcal O(M)$, providing advantages for scalability to large networks. Additionally, information from end times is added to the CIF, providing a more complete modelling framework for event data on networks with an associated event duration.

\section{Graph-based mutually exciting processes} \label{sec:gbmepp}

The main objective of this work is to propose a graph-based mutually exciting model (GB-MEP) for the node-specific conditional intensity functions $\lambda_i(t)$ for the temporal point processes of start times of connection events on a graph, applied to journeys observed on bike-sharing systems.
The GB-MEP intensity function is assumed to take the following generic form:
\begin{equation}
\lambda_i(t) = \lambda_i + \sum_{j=1}^M \left\{\sum_{k=1}^{N_j(t)}\omega_i\{\gamma(i,j), t-t_{j,k}\} + \sum_{h=1}^{N_j^\prime(t)}\omega_i^\prime\{\gamma(i,j), t-t^\prime_{j,h}\}\right\},
\label{cif_full}
\end{equation}
where $\lambda_i\in\mathbb R_+$ is a baseline rate, $\omega_i,\omega_i^\prime: \mathbb R_+\times\mathbb R_+\to\mathbb R_+$ are spatio-temporal excitation kernel functions, and $\gamma:\mathcal V\times\mathcal V\to\mathbb R_+$ is a distance function between nodes. 
The distance function $\gamma$ quantifies how close in space two nodes are. For example, in bike-sharing systems, the distance function could be taken to be the haversine metric, which provides an \textit{``as-the-crow-flies''} measure of distance between the geographic location of two nodes. Given the latitude and longitude $(\varphi_x,\ell_x)$ and $(\varphi_y,\ell_y)$ of the bike stations $x$ and $y$ (in radians), the haversine distance between the stations is:
\begin{equation}
\gamma(x,y) = 2\rho\arcsin\left\{\sqrt{\sin^2\left(\frac{\varphi_y-\varphi_x}{2}\right) + \cos(\varphi_x)\cos(\varphi_y)\sin^2\left(\frac{\ell_y-\ell_x}{2}\right)} \right\}, \label{dist}
\end{equation}
where $\rho$ is the Earth radius. Note that, for more general point processes on networks where information about the node locations is not present, the distance $\gamma(x,y)$ between two nodes $x,y\in\mathcal V$ could depend on graph-related statistics, such as the length of the shortest path between the nodes, or the number of common neighbours. In this case $\gamma$ takes the interpretation of a \textit{dissimilarity metric}. 

\begin{figure}[!t]
\centering
\centering
\includegraphics[width=0.9\textwidth]{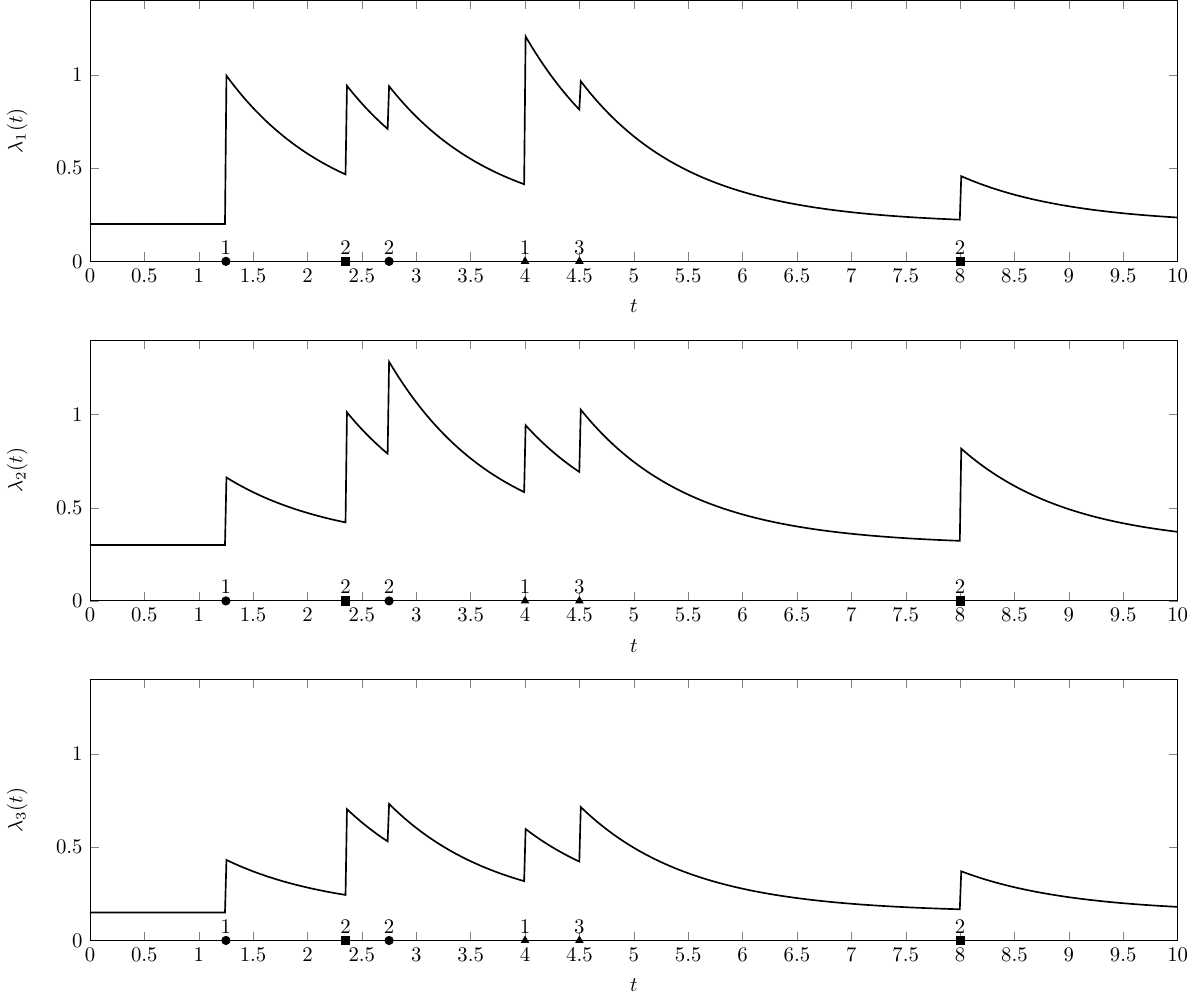}
\caption{Example of CIFs for a GB-MEP with $\mathcal V=\{1,2,3\}$ and three realised events expressed as quadruples $(x,y,t,t^\prime),\ x,y\in\mathcal V,\ t,t^\prime\in\mathbb R_+,\ t<t^\prime$. Events and their symbols: $(1,2,1.25,2.75)$ corresponds to $\bullet$; $(1,3,4,4.5)$ to $\blacktriangle$, and $(2,2,2.35,8)$ to $\blacksquare$. Parameters: $\lambda_1=0.2$, $\lambda_2=0.3$, $\lambda_3=0.15$, $\alpha_1=0.8$, $\alpha_2=\alpha_3=0.6$, $\beta_1=\beta_2=\beta_3=1$, $\theta_1=\theta_2=\theta_3=1$, $\alpha_1^\prime=\alpha_2^\prime=0.5$, $\alpha_3^\prime=0.3$, $\beta_1^\prime=\beta_2^\prime=\beta_3^\prime=1$, $\theta_1^\prime=\theta_2^\prime=\theta_3^\prime=1.5$. Node distances: $\gamma(1,2)=\gamma(2,1)=0.5$, $\gamma(1,3)=\gamma(3,1)=0.75$, $\gamma(2,3)=\gamma(3,2)=0.25$, $\gamma(1,1)=\gamma(2,2)=\gamma(3,3)=0$.}
\label{fig:example}
\end{figure}

In this work, it is further assumed that the spatial and temporal components of the excitation kernels $\omega_i$ are separable, with the temporal element of the kernel taking a scaled exponential form. This results in the following excitation kernel:
\begin{equation}
\omega_i(\gamma^\ast, t) = \kappa_i(\gamma^\ast)\cdot\alpha_i\exp(-\beta_i t), 
\label{exc_ker}
\end{equation}
where $\alpha_i,\beta_i\in\mathbb R_+$, $\alpha_i<\beta_i$, and $\gamma^\ast$ corresponds to a realised value of the distance function $\gamma$. The function $\kappa_i:\mathbb R_+\to(0,1]$ rescales the node-specific temporal mutual excitation given by $\alpha_i\exp(-\beta_i t)$, using the distance $\gamma$ between the nodes. 
In the application to bike-sharing systems, the excitation effect of an event occurring on a node with small distance from station $i$ is likely to have a larger effect on the intensity $\lambda_i(t)$ than an event occurring on a node located further away. Therefore, $\kappa_i$ could be set to a monotonically decreasing function such as $\kappa(0)=1$ and $\lim_{\gamma^\ast\to\infty}\kappa_i(\gamma^\ast)=0$. A possible choice, used in this work, is the truncated exponential function $\kappa_i(\gamma^\ast)=\exp(-\theta_i\gamma^\ast)\mathds 1_{[0,\varepsilon]}(\gamma^\ast)$, where $\theta_i\in\mathbb R_+$ is a decay parameter and $\varepsilon\in\mathbb R_+$ is a \textit{fixed constant} defining the radius of neighbouring nodes considered to have non-zero effect on the CIF $\lambda_i(t)$. Note that setting $\varepsilon\to0$ would result in $\kappa_i(\gamma^\ast)=\mathds 1_{\{0\}}(\gamma^\ast)$, which would only give weight to events involving station $i$, removing the spatial mutual excitation component, and making the process exclusively self-exciting. 
A similar spatio-temporal excitation kernel could also be postulated for $\omega^\prime_i(\gamma^\ast, t)$:
\begin{equation}
\omega_i^\prime(\gamma^\ast, t) = \exp(-\theta_i^\prime\gamma^\ast)\mathds 1_{[0,\varepsilon]}\{\gamma^\ast\}\cdot\alpha_i^\prime\exp(-\beta_i^\prime t), 
\end{equation}
with $\alpha_i^\prime,\beta_i^\prime,\theta_i^\prime
\in\mathbb R_+$ and $\alpha_i^\prime<\beta_i^\prime$. 
Overall, this gives the following extended form for the proposed GB-MEP CIF in Equation~\eqref{cif_full} for applications to BSSs: 
\begin{multline}
\lambda_i(t)=\lambda_i+\sum_{j=1}^M \Bigg\{
\sum_{k=1}^{N_j(t)} 
\exp\{-\theta_i\gamma(i,j)\}\mathds 1_{[0,\varepsilon]}\{\gamma(i,j)\}
\cdot\alpha_i\exp\{-\beta_i (t - t_{j,k})\} \\ +
\sum_{h=1}^{N_j^\prime(t)} \exp\{-\theta_i^\prime\gamma(i,j)\} \mathds 1_{[0,\varepsilon]}\{\gamma(i,j)\} \cdot \alpha^\prime_i\exp\{-\beta_i^\prime(t-t_{j,h}^\prime)\}\Bigg\}. 
\label{full_cif}
\end{multline}
An example of the intensity functions arising under a GB-MEP process is given in Figure~\ref{fig:example}.

\subsection{Parameter estimation} \label{sec:parest}

The model parameters $\boldsymbol\phi_i=(\lambda_i,\alpha_i,\beta_i,\theta_i,\alpha_i^\prime,\beta_i^\prime, \theta_i^\prime)$ for each node-based process could be learned via maximum likelihood estimation, by maximising the node-specific log-likelihood function separately for each station. The log-likelihood function for a temporal point process takes the following form \citep[see, for example,][]{Daley02}:
\begin{align}
\ell_i(\boldsymbol\phi_i) = \sum_{k=1}^{N_i(T)}\log[\lambda_i(t_{i,k})] - \Lambda_i(T), \label{loglik}
\end{align}
where $\Lambda_i:\mathbb R_+\to\mathbb R_+$ is the compensator, defined as 
\begin{align}
\Lambda_i(t) = \int_{0}^t\lambda_i(s)\mathrm ds. 
\end{align}
For the CIF in Equation~\eqref{full_cif}, the compensator takes the following form:
\begin{multline}
\Lambda_i(t) = \lambda_i t\ - \sum_{j=1}^M \Bigg\{\frac{\exp\{-\theta_i\gamma(i,j)\}\mathds 1_{[0,\varepsilon]}\{\gamma(i,j)\} \cdot \alpha_i}{\beta_i} \sum_{k=1}^{N_j(t)}[\exp \{-\beta_i(t-t_{j,k})\}-1] \\ - 
\frac{\exp\{-\theta_i^\prime\gamma(i,j)\} \mathds 1_{[0,\varepsilon]}\{\gamma(i,j)\} \cdot\alpha_i^{\prime}}{\beta_i^{\prime}} \sum_{h=1}^{N_i^{\prime}(t)}[\exp \{-\beta_i^{\prime}(t-t_{j,h}^{\prime})\}-1]\Bigg\}.
\end{multline}
Note that, because of the double summations in the CIF in Equation~\eqref{full_cif}, the computational complexity for calculating the term $\sum_{k=1}^{N_i(T)}\log[\lambda_i(t_{i,k})]$ in 
Equation~\eqref{loglik} is $\mathcal O\{N_i(T)\sum_{j=1}^M \mathds 1_{[0,\varepsilon]}\{\gamma(i,j)\}[N_j(T) + N_j^\prime(T)]\}$, proportional to the sum of the number of events occurring on each node, multiplied by the total number of events started from the station under consideration.
In practice, this makes model fitting particularly challenging, and almost unfeasible for large datasets.  
This issue is largely alleviated by the scaled exponential excitation kernel, which admits a computationally cheaper form of the log-likelihood, written in recursive form using the following expression for the CIF in Equation~\eqref{full_cif}:
\begin{equation}
\lambda_i(t_{i,k})=\lambda_i +
\sum_{j=1}^M \left[ \exp\{-\theta_i\gamma(i,j)\}\mathds 1_{[0,\varepsilon]}\{\gamma(i,j)\} \cdot\alpha_i A_{ij}(k) +
\exp\{-\theta_i^\prime\gamma(i,j)\} \mathds 1_{[0,\varepsilon]}\{\gamma(i,j)\} \cdot\alpha^\prime_i A_{ij}^\prime(k) \right]. 
\end{equation}
In the above expression, $A_{ij}(k)$ and $A_{ij}^\prime(k)$ are defined as follows: 
\begin{align}
A_{ij}(k) = \sum_{l=1}^{N_j(t_{i,k})} \exp\{-\beta_i(t_{i,k}-t_{j,l})\}, & & 
A_{ij}^\prime(k) = \sum_{l=1}^{N_j^\prime(t_{i,k})} \exp\{-\beta_i^\prime(t_{i,k}-t_{j,l}^\prime)\}. \label{eq:A}
\end{align}
The initial values $A_{ij}(1)$ and $A^\prime_{ij}(1)$ are simply:
\begin{align}
A_{ij}(1)=\sum_{l=1}^{N_j(t_{i,1})} \exp\{-\beta_i(t_{i,1}-t_{j,l})\}, & &
A^\prime_{ij}(1)=\sum_{l=1}^{N_j^{\prime}(t_{i, 1})} \exp\{-\beta_i^\prime(t_{i, 1}-t_{j, l}^{\prime})\}.
\end{align}
The recursive form of the log-likelihood stems from the following proposition.

\begin{proposition} \label{prop:A}
The expressions for 
$A_{ij}(k)$ and $A_{ij}^\prime(k)$ in Equation~\eqref{eq:A} can be rewritten recursively as follows: 
\begin{align}
& A_{ij}(k)=\exp\{-\beta_i(t_{i, k}-t_{i, k-1})\} \cdot A_{ij}(k-1) + \sum_{l=N_j(t_{i,k-1})+1}^{N_j(t_{i,k})} \exp\{-\beta_i(t_{i, k}-t_{j, l})\}, \label{eq:A_rec} \\
& A_{ij}^{\prime}(k)=\exp\{-\beta_i^{\prime}(t_{i, k}-t_{i, k-1})\} \cdot A_{i j}^{\prime}(k-1) +\sum_{l=N_j^{\prime}(t_{i, k-1})+1}^{N_j^{\prime}(t_{i, k})} \exp\{-\beta_i^\prime(t_{i, k}-t_{j, l}^{\prime})\}.
\end{align}
\end{proposition}\label{prop:rec}
\begin{proof}
The result is proved for $A_{ij}(k)$ here; the proof for $A_{ij}^{\prime}(k)$ is analogous. The summation for $A_{ij}(k)$ in Equation~\eqref{eq:A} could be broken up at $N_j(t_{i,k-1})$ as follows: 
\begin{align}
A_{ij}(k) = \sum_{l=1}^{N_j(t_{i,k-1})} \exp\{-\beta_i(t_{i,k}-t_{j,l})\} + \sum_{l=N_j(t_{i,k-1})+1}^{N_j(t_{i,k})} \exp\{-\beta_i(t_{i,k}-t_{j,l})\}. \label{eq:A_broken}
\end{align}
Pre-multiplying the first element of Equation~\eqref{eq:A_broken} by $\exp\{-\beta_i(t_{i,k-1})-\beta_i(t_{i,k-1})\}$ gives:
\begin{align}
\sum_{l=1}^{N_j(t_{i,k-1})} \exp\{-\beta_i(t_{i,k}-t_{j,l})\} &= \exp\{-\beta_i(t_{i,k-1})-\beta_i(t_{i,k-1})\}\sum_{l=1}^{N_j(t_{i,k-1})} \exp\{-\beta_i(t_{i,k}-t_{j,l})\} \\ &= \exp\{-\beta_i(t_{i,k})-\beta_i(t_{i,k-1})\}\sum_{l=1}^{N_j(t_{i,k-1})} \exp\{-\beta_i(t_{i,k-1}-t_{j,l})\}. 
\end{align} 
By the definition in Equation~\eqref{eq:A}, $\sum_{l=1}^{N_j(t_{i,k-1})} \exp\{-\beta_i(t_{i,k-1}-t_{j,l})\} = A_{ij}(k-1)$, which gives the result in Equation~\eqref{eq:A_rec}.
\end{proof}

It follows that, writing $B_{ij}(k)=
\exp\{-\theta_i\gamma(i,j)\}\mathds 1_{[0,\varepsilon]}\{\gamma(i,j)\} \cdot \alpha_i A_{ij}(k) + \exp\{-\theta_i^\prime\gamma(i,j)\} \mathds 1_{[0,\varepsilon]}\{\gamma(i,j)\} \cdot\alpha_i^\prime A_{ij}^\prime(k)$, 
Proposition~\ref{prop:A} gives a recursive form for the log-likelihood in Equation~\eqref{loglik}:
\begin{align}
\ell_i(\boldsymbol\phi_i) = \sum_{k=1}^{N_i(T)}\log\left\{\lambda_i+ \sum_{j=1}^M B_{ij}(k) \right\} - \Lambda_i(T). 
\label{eq:recursive_loglik}
\end{align}
Under this recursive representation, $\ell_i(\boldsymbol\phi_i)$ can be evaluated in $\mathcal O\{N_i(T)M_i^{\varepsilon}\}$, where $M_i^{\varepsilon}=\sum_{j=1}^M\mathds 1_{[0,\varepsilon]}\{\gamma(i,j)\}$ is the number of nodes within distance $\varepsilon$ of node $i$. Note that this assumes that the update of $B_{ij}(k)$ given $B_{ij}(k-1)$ is $\mathcal O(M_i^{\varepsilon})$. 

The node-specific log-likelihoods $\ell_i(\boldsymbol\phi_i),\ i=1,\dots,M$ can be optimised independently and in parallel via standard statistical optimisation techniques. For example, the node-based expectation-maximisation (EM) algorithm described in \cite{Fox16} can be adapted to GB-MEP utilising the same latent variables and reparametrisation discussed in \cite{Fox16} and \cite{SannaPassino23}. The main drawback of optimising the log-likelihood via the EM algorithm consists in its computation cost: the algorithm augments the parameter space with latent variables used to reconstruct the branching structure associated with the process \citep[see, for example, Section 3.3 in][]{Laub22}. Therefore, it is computationally more convenient to use gradient-based methods for optimisation, leveraging the recursive structure of the log-likelihood stemming from Proposition~\ref{prop:A}. In particular, this work uses the limited-memory Broyden-Fletcher-Goldfarb-Shanno (L-BFGS) algorithm for optimisation \citep[see, for example,][]{Zhu97}, a quasi-Newton method which approximates the inverse of the Hessian matrix using gradients. Using the form of the log-likelihood in Equation~\eqref{eq:recursive_loglik}, recursive expressions for the gradients of $\ell_i(\boldsymbol\phi_i)$ can be obtained with respect to all parameters in $\boldsymbol\phi_i$, implying that estimation via L-BFGS is computationally feasible even for large datasets. 

\subsection{Model evaluation}

The quality of the model fit could be assessed by evaluating the distribution of $p$-values calculated on a training and a test set observation periods, using the same methodology discussed in \cite{SannaPassino23}.
By the time rescaling theorem \citep[see, for example,][]{Brown02}, the transformed event times $\Lambda_i(t_{i,k}),\ k=1,2,\dots$
are event times of a homogeneous Poisson process with unit rate, under the null hypothesis of correct specification of the conditional intensity and its parameters. Therefore, under the same null hypothesis, the upper tail $p$-values calculated from the observed events should follow a uniform distribution in $(0,1)$. Since the inter-arrival times of a unit rate Poisson process are exponentially distributed with unit rate, the corresponding upper tail $p$-values for the observed events are:
\begin{equation}
p_{i,k} = \exp\left\{-\int_{t_{i,k-1}}^{t_{i,k}} \lambda_i(s)\mathrm ds\right\} = \exp\{-\Lambda_i(t_{i,k}) + \Lambda_i(t_{i,k-1})\},\ i=1,\dots,M,\ k=1,\dots,N_i(T), \label{eq:pval}
\end{equation}
assuming $\Lambda_i(t_{i,0})=0$ for all $i=1,\dots,M$.
Using the recursive expressions in Proposition~\ref{prop:rec} and the counting processes in Equation~\eqref{eq:counting_simple}, the expression in Equation~\eqref{eq:pval} reduces to:
\begin{multline}
p_{i,k} = \exp\Bigg\{ \sum_{j=1}^M \frac{\exp\{-\theta_i\gamma(i,j)\}\mathds 1_{[0,\varepsilon]}\{\gamma(i,j)\}  \cdot \alpha_i}{\beta_i} [A_{ij}(k)-A_{ij}(k-1)-N_j(t_{i,k})+N_j(t_{i,k-1})]  \\ 
 +\sum_{j=1}^M \frac{\exp\{-\theta_i^\prime\gamma(i,j)\}\mathds 1_{[0,\varepsilon]}\{\gamma(i,j)\}  \cdot \alpha_i^{\prime}}{\beta_i^{\prime}} [A_{ij}^\prime(k)-A_{ij}^\prime(k-1)-N_j^\prime(t_{i,k})+N_j^\prime(t_{i,k-1})] 
-\lambda_i(t_{i,k}-t_{i,k-1})
\Bigg\}.
\end{multline}
Considering an observation period $[0,T]$, the model parameters could first be estimated from a testing period $[0,T^\ast]$, $T^\ast<T$, using the procedure described in Section~\ref{sec:parest}. The $p$-values can then be calculated directly by plugging in the estimates into Equation~\eqref{eq:pval}. The estimates could also be used to evaluate the performance of the model in a testing period $(T^\ast,T]$, by calculating the $p$-values using the same methodology. In both cases, under a correct model specification and parameter estimation, the distributions of the $p$-value on the training and test set events should be approximately uniform in $(0,1)$. 
The quality of the fit to the uniform distribution could be evaluated using 
the Kolmorogov-Smirnov statistic, denoted here $\mathrm{KS}_i$. 
For the sequence of node-specific $p$-values $p_{i,k},\ k=1,\dots,N_i(T)$ on the training set, 
 $\mathrm{KS}_i$ is defined as follows:
\begin{equation}
\mathrm{KS}_i  = \sup_{p\in[0,1]} \left\vert\ p - \frac{1}{N_i(T)}\sum_{k=1}^{N_i(T)} \mathds 1_{[0,p)}(p_{i,k})\ \right\vert. 
\end{equation}

\section{Application to the Santander Cycles bike-sharing system} \label{sec:results}

The proposed models were tested on data from the Santander Cycles bike-sharing network in central London. Transport for London publicly released data about journeys on the Santander Cycles network, publicly available at \url{https://cycling.data.tfl.gov.uk/} (powered by TfL). Each record includes information about the start and end times, IDs of source and destination stations, journey duration, a bike ID number, and an ID number for the rental. 
For the analysis in this section, 32 weeks of data were considered, starting from 2\textsuperscript{nd} March 2022 until 9\textsuperscript{th} October 2022. A total of $\numprint{4098756}$ journeys recorded between 2\textsuperscript{nd} March 2022 and 21\textsuperscript{st} June 2022 (approximately 16 weeks) were used to construct the training set, whereas the remaining $\numprint{3995591}$ journeys occurred between 22\textsuperscript{nd} June 2022 and 9\textsuperscript{th} October 2022 (approximately 16 weeks)  formed the test set. A total of $M=787$ unique active docking stations were observed in the data under consideration. Distances between docking stations were calculated using the haversine distance formula in Equation~\eqref{dist}, with the value $\rho=6365.079\mathrm{km}$ (corresponding to the Earth radius at the latitude of London, with no altitude). A more precise estimate of the cycling distance between two nodes could be obtained via calls to the TfL Journey Planner API, or any other tool providing calculations of cycling distances between locations. The main practical limitation for this approach is that the calculation of the full distance matrix would require ${M\choose{2}}$ API requests, which normally largely exceeds the number of permitted requests under a basic plan.

The proposed GB-MEP model in Equation~\ref{cif} has been compared to a number of alternatives:
\begin{itemize}
\item A standard Poisson process model for each station (Poisson): $$\lambda_i(t)=\lambda_i.$$
\item A mutually exciting model for each station (MEP): $$\lambda_i(t)=\lambda_i+\sum_{h=1}^{N_i^\prime(t)} \alpha_i^\prime\exp\{-\beta_i^\prime(t-t_{i,h}^\prime)\}.$$
\item A self-exciting Hawkes process model for each station (SEP): $$\lambda_i(t)=\lambda_i+\sum_{k=1}^{N_i(t)} \alpha_i\exp\{-\beta_i^\prime(t-t_{i,k})\}.$$
\item A self-and-mutually exciting process for each station (SMEP): $$\lambda_i(t)=\lambda_i+\sum_{k=1}^{N_i(t)} \alpha_i\exp\{-\beta_i^\prime(t-t_{i,k})\}+\sum_{h=1}^{N_i^\prime(t)} \alpha_i^\prime\exp\{-\beta_i^\prime(t-t_{i,h}^\prime)\}.$$
\item A spatial mutually exciting process for each station (SpMEP): $$\lambda_i(t)=\lambda_i+\sum_{j=1}^M \sum_{k=1}^{N_j(t)} \exp\{-\theta_i\gamma(i,j)\}\mathds 1_{[0,\varepsilon]}\{\gamma(i,j)\}\cdot\alpha_i\exp\{-\beta_i (t - t_{j,k})\}.$$
\end{itemize}
Note that all models above can be written as special cases of GB-MEP for specific choices of the functions $\kappa$ and $\kappa^\prime$, \textit{cf.} Equation~\eqref{exc_ker}. 
The value of $\varepsilon$ in SpMEP and GB-MEP has been set to $\varepsilon=0.5$. For some stations, $M_i^\varepsilon<3$ for $\varepsilon=0.5$; in these cases, $\varepsilon$ was set to the minimum value such that $M_i^\varepsilon\geq 3$. 
All model parameters for GB-MEP and its competitors were estimated via maximum likelihood, using numerical optimisation via L-BFGS. Note that the Poisson model has a closed form solution: $\hat\lambda_i=N_i(T^\ast)/T^\ast$ for a training period $[0,T^\ast)$. The parameters of SEP and MEP were initialised to $(\lambda_i,\alpha_i,\beta_i)=(e^{-4},e^{-4},2e^{-4})$, and the corresponding results were used to initialise SMEP. In turn, the results obtained from SMEP were used to initialise SpMEP and GB-MEP.  

\begin{figure}[!t]
\centering
\begin{subfigure}[t]{.475\textwidth}
\centering
\caption{Training set}
\label{fig:qq_train}
\includegraphics[width=\textwidth]{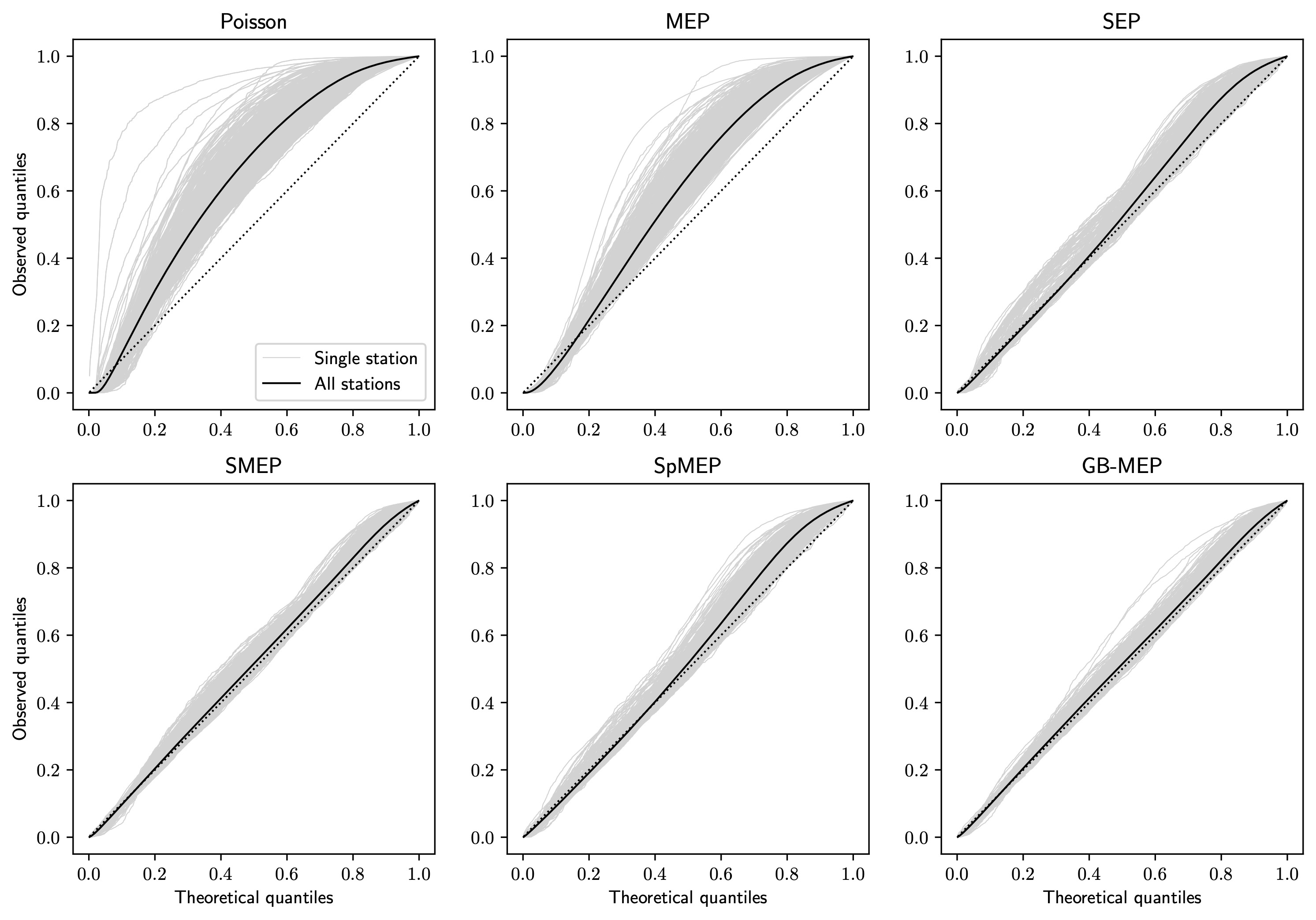}
\end{subfigure}
\unskip\ \vrule\ 
\begin{subfigure}[t]{.475\textwidth}
\centering
\caption{Test set}
\label{fig:qq_test}
\includegraphics[width=\textwidth]{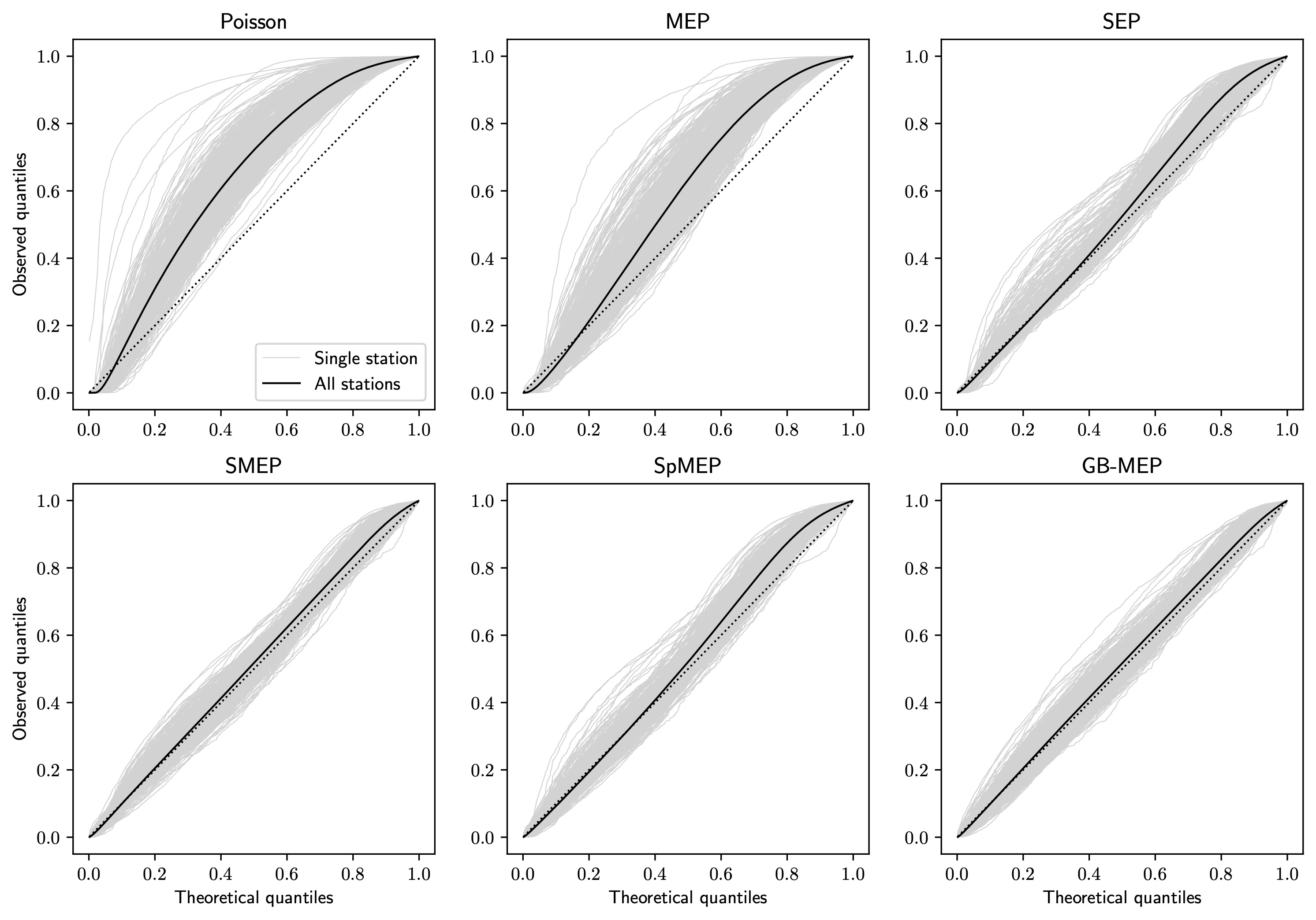}
\end{subfigure}
\caption{Uniform Q-Q plots of $p$-values, calculated separately for each station on the training and test sets for six models.}
\label{fig:qq}
\end{figure}

\begin{figure}[!t]
\centering
\begin{subfigure}[t]{.475\textwidth}
\centering
\caption{Training set}
\label{fig:box_train}
\includegraphics[width=\textwidth]{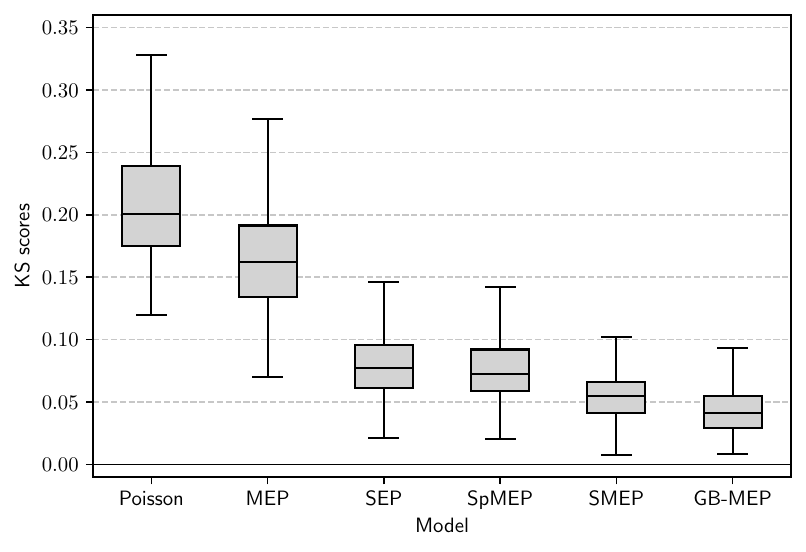}
\end{subfigure}
\unskip\ \vrule\ 
\begin{subfigure}[t]{.475\textwidth}
\centering
\caption{Test set}
\label{fig:box_test}
\includegraphics[width=\textwidth]{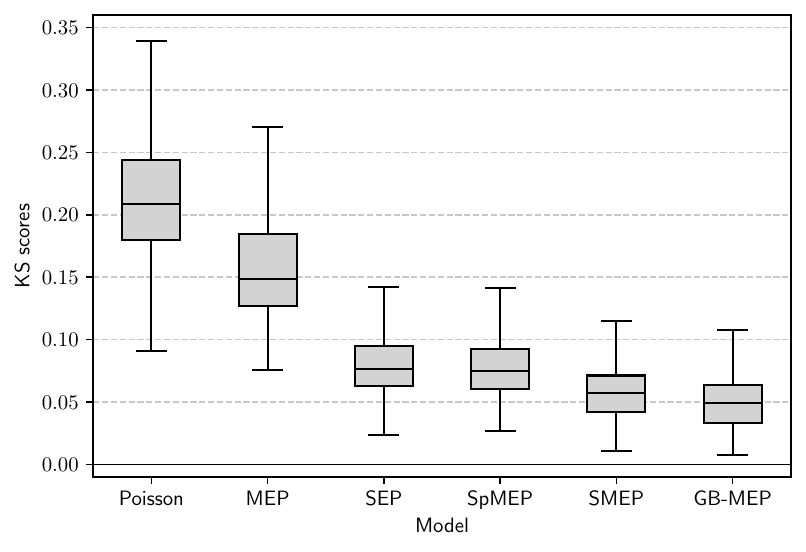}
\end{subfigure}
\caption{Boxplots of KS scores across all station, calculated on the training and test sets for six models.}
\label{fig:box}
\end{figure}

The results are reported in Table~\ref{tab:ks}, Figure~\ref{fig:qq} and Figure~\ref{fig:box}. Table~\ref{tab:ks} shows the Kolmogorov-Smirnov score for the distribution of $p$-values across \textit{all $M$ stations}, calculated against the uniform distribution. Figure~\ref{fig:qq} reports uniform quantile-quantile (Q-Q) plots for all stations individually, and for all stations combined. Figure~\ref{fig:box} contains the boxplots of KS scores for all six models considered in this section. The results show that GB-MEP has the best performance across all models, both on the training and test set, demonstrating that taking into account the geographical location of nodes within the network is important to improve the goodness-of-fit of statistical models to real-world network point process data. GB-MEP is the only model which achieves a median KS score below 0.05 across stations, both in the training and test sets (\textit{cf.} Figure~\ref{fig:box}). Additionally, Table~\ref{tab:ks} shows remarkable agreement between the results on the training and test sets, which suggests that the models are not overfitting the data. 

\begin{table}[t]%
\centering
\caption{Kolmogorov-Smirnov scores for the distribution of $p$-values for all stations on training and test sets for six models.\label{tab:ks}}%
\begin{tabular}{lcccccc}%
\toprule
& Poisson & MEP & SEP & SMEP & SpMEP & GB-MEP \\
\midrule
Training set & $0.2186$ & $0.1610$ & $0.0751$ & $0.0326$ & $0.0734$ & $0.0246$ \\
Test set & $0.2217$ & $0.1584$ & $0.0752$ & $0.0350$ & $0.0738$ & $0.0264$ \\
\bottomrule
\end{tabular}
\end{table}

\begin{figure}[!t]
\centering
\begin{subfigure}[t]{.545\textwidth}
\centering
\caption{Plot of station locations, coloured according to the difference in KS scores between SMEP and GB-MEP on the test set}
\label{fig:london}
\includegraphics[width=\textwidth]{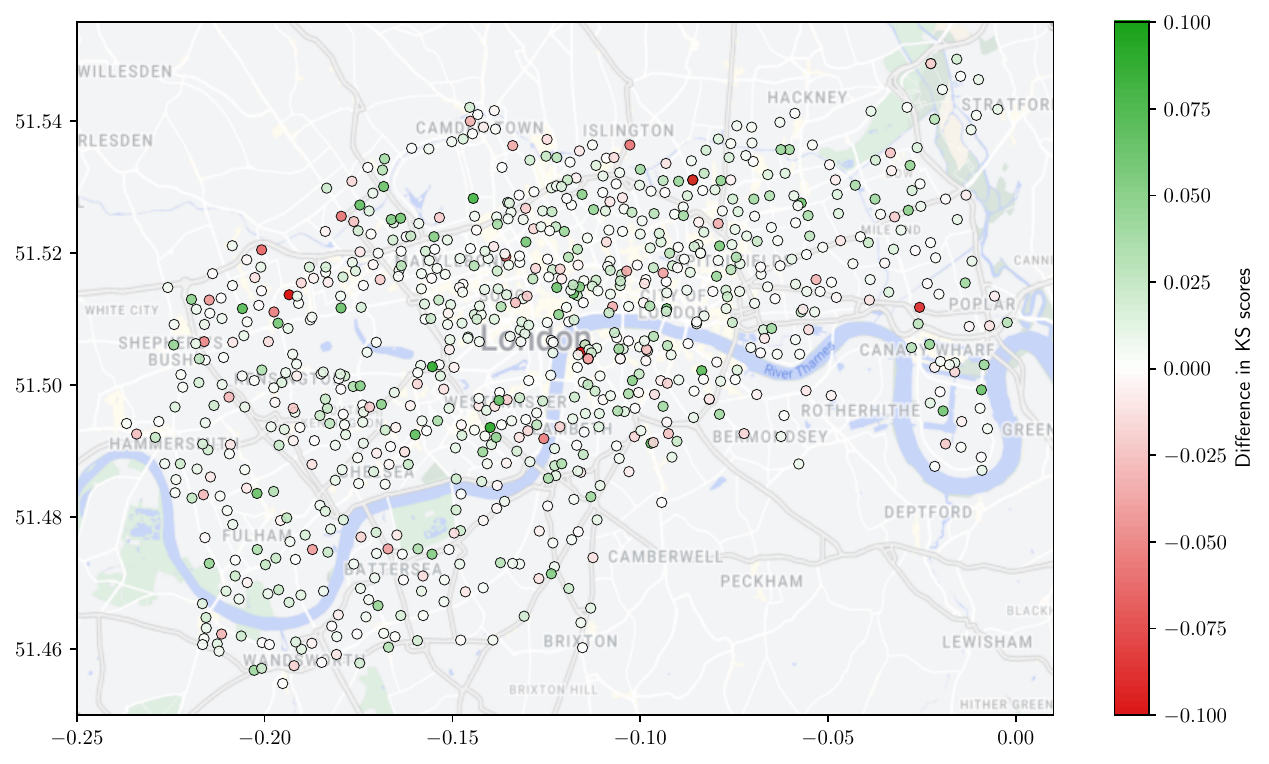}
\end{subfigure}
\unskip\ \vrule\ 
\begin{subfigure}[t]{.435\textwidth}
\centering
\caption{Histogram of differences between KS scores for SMEP and GB-MEP on the test set}
\label{fig:hist}
\includegraphics[width=0.97\textwidth]{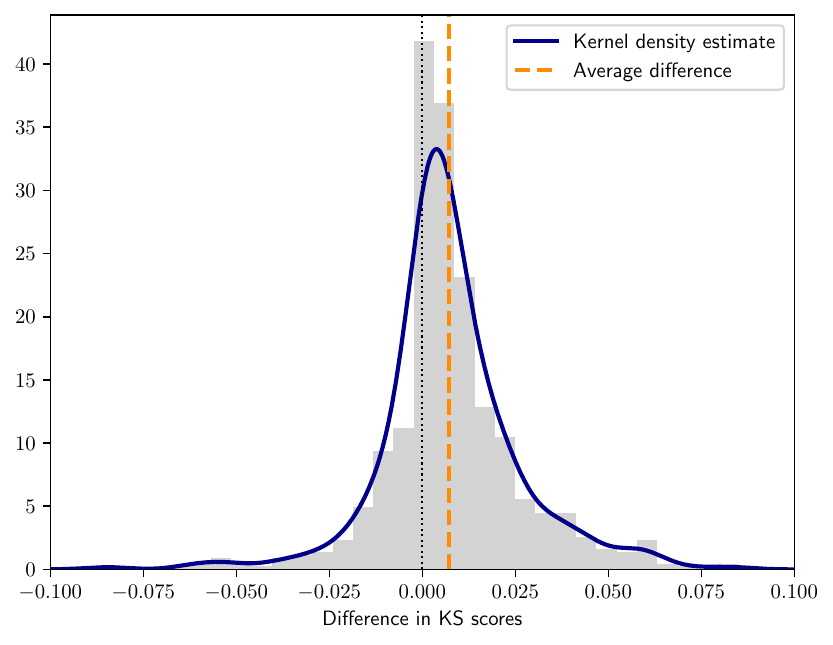}
\end{subfigure}
\caption{Station locations and histogram 
of differences in KS scores between SMEP and GB-MEP on the test set.}
\label{fig:differences}
\end{figure}

For most stations, the two best performing models are SMEP and GB-MEP, as observed in Table~\ref{tab:ks}, Figure~\ref{fig:qq} and Figure~\ref{fig:box}. GB-MEP has a better performance than SMEP on approximately 73.88\% of stations in the test set. Figure~\ref{fig:differences} analyses more closely the differences between the results of GB-MEP and SMEP, showing all docking stations in London, coloured by the difference in KS scores on the test set, and a histogram of differences between KS scores for SMEP and GB-MEP on the test set. 

The results demonstrate that GB-MEP is an appropriate choice for modelling bike-sharing systems. The spatial component in GB-MEP implies that events are not only influenced by past occurrences, but they also depend on the spatial proximity to other locations. This means that high demand for bikes in one station can influence neighbouring stations. For example, all docking stations around railway stations in central London periodically witness increased demand from commuters seeking transportation to their workplaces during morning rush hour. As one moves further away from the railway stations, the demand for bikes diminishes, mainly due to decreased proximity and preferences of commuters. This behaviour is easily captured by GB-MEP. 

In addition, GB-MEP can also capture increased demand due to non-periodic events. For example, consider a sporting event or a concert in central London, which often result in concentrated demands for transportation. As fans leave the stadium or concert hall, there would likely be a surge in demand for bikes at the docking stations close to the event location. Nearby stations, especially those within a reasonable walking distance from the stadium or concert hall, would likely experience increased demand as well.
The spatial component in GB-MEP explicitly accounts for the spatial proximity of bike stations, resulting in a particularly suitable model for situations when one station experiences a high demand for bikes due to a specific event or circumstance, triggering a chain reaction affecting nearby stations as people move around the area.

The results also show an effect of the end times on the intensity of start times of journeys, which may seem counterintuitive, but it is consistent with some characteristics of the Santander Cycles BSS: for example, if a station is empty, a user returning a bike to that station may imply that there is a higher chance of a journey starting immediately from that station, with a different user picking up the newly available bike. 
Also, until September 2022, users were charged \textsterling2 for unlimited journeys with Santander Cycles in the subsequent 24 hours. For journeys longer than 30 minutes, users are additionally charged. Therefore, many users typically returned bikes before the 30 minutes mark, and they would pick up another bike immediately for a new journey, meaning that returning a bike to a station could increase the likelihood of a new journey starting from the same station.

\section{Conclusion} \label{sec:discussion}

This article discussed a general strategy to incorporate measures of distance between nodes on a graph within the context of multivariate Hawkes processes on graphs, proposing GB-MEP, a graph-based mutually exciting point process. Scalable inference procedures were discussed, using a recursive form of the likelihood function under a scaled exponential form of the excitation kernel. The model was fitted on a large dataset of journeys within the Santander Cycles bike-sharing system, with over 4 million events in the training set, and close to $M=800$ nodes. The model performance was evaluated on close to 4 million events comprising the test set, demonstrating superior performance of GB-MEP over alternative 
approaches. 

A possible limitation of the proposed framework for the application to bike-sharing systems is that a bound on the capacity of each station is not included within the model: each station has a limited number of docks, which could be used to correct the CIF when a station is at capacity. Also, additional distance functions could be considered for the excitation component.
This work is framed within the context of docked bike-sharing systems, but it could be more generally applied to mutually exciting processes on any type of network where a distance metric between nodes is available. If an adjacency matrix is observed, and not a distance matrix, measures of similarity between nodes could be used for $\gamma$ within the GB-MEP framework (for example, the length of the shortest path between nodes).  

A possible extension of the version of GB-MEP presented in this work would be to consider different excitation kernels, such as power laws \citep[see, for example,][]{Laub22}. Furthermore, removing the separability assumption on the excitation kernel could also prove interesting, as it would allow to capture more complex patterns arising from the interaction between the spatial and temporal components. 
Finally, adding a probabilistic mixture model component in the bike-sharing application could further improve the model 
performance, reflecting the fact that bikes could be used for different purposes, such as leisure or commuting. 

\section*{Code}

A \textit{Python} library with code and examples to reproduce all results in this article can be found in the GitHub repository \href{https://github.com/fraspass/gb-mep}{\texttt{fraspass/gb-mep}}. The README file in the repository provides detailed instructions on how to download the Santander Cycles data, preprocess them, and run the experiments.


\bibliographystyle{rss}
\singlespacing
\bibliography{biblio}

\end{document}